# Site-selective synthesis of *in situ* Ni-filled multi-walled carbon nanotubes using Ni(salen) as a catalyst source

**Joydip Sengupta**[1], **Avijit Jana**[2], **N D Pradeep Singh**[2], **C Mitra**[3] **and Chacko Jacob**[1,4]

[1] Materials Science Centre, Indian Institute of Technology, Kharagpur 721302, India
[2] Department of Chemistry, Indian Institute of Technology, Kharagpur 721302, India
[3] Department of Physical Science, Indian Institute of Science Education and Research, Kolkata 741252, India

E-mail: cxj14_holiday@yahoo.com

**Abstract**
The synthesis of Ni-filled multi-walled carbon nanotubes was performed by atmospheric pressure chemical vapor deposition with propane on Si at 850 °C using a simple mixture of ($N, N'$-bis(salicylidene)-ethylenediiminato) nickel(II), commonly known as Ni(salen), and a conventional photoresist. Analysis of the carbon nanotubes using scanning electron microscopy together with high-resolution transmission electron microscopy show that the nanotubes have grown by a tip-growth mechanism and exhibit a multi-walled structure with partial Ni filling. The high quality of the Ni-filled nanotubes is evidenced by Raman spectroscopy. The magnetic properties of Ni-filled nanotubes were analyzed using a superconducting quantum interference device which revealed their ferromagnetic behavior with large coercivity. A scalable as well as site-selective growth of high quality Ni-filled carbon nanotubes is achieved by a simple photolithographic method.

## 1. Introduction

In recent years, magnetic-material-filled carbon nanotubes (CNTs) have been very attractive to researchers due to their potential application in magnetic force microscopy [1], high density magnetic recording media [2], biology [3, 4], microwave absorption [5], drug delivery [6], molecular spintronics [7], nano-electromagnetic inductors [8], nanopipettes [9], nanowelding [10], electrochemical energy storage [11] and even for cancer treatment [12]. Theoretical studies [13, 14] have already suggested that incorporation of nanoparticles of foreign materials into the cavity of the nanotubes may significantly modify the electronic and mechanical properties of the nanotubes, as well as alter the properties of the filling materials. The CNT-encapsulated nanoparticles have excellent thermal and chemical stabilities as they are well protected by the CNT against coarsening and oxidation, which promote them as a potential candidate for application in harsh industrial environments where the stability of the nanostructures is crucial. Different methods of filled CNT synthesis including capillary infiltration [15], chemical methods [16], arc-discharge [17], electrolysis of molten salt [18], electrochemical deposition [19] and CVD [20–23] have already been reported by several research groups. From a technological viewpoint, besides the mere growth of metal-filled CNTs, the exact positioning of those filled nanotubes over the desired substrate is a major challenge for integration of filled nanotubes in a device structure. However, despite tremendous progress in synthesizing CNTs, the effective way to directly obtain site-selective growth of *in situ* metal-filled carbon nanotubes over the pre-defined location has not been extensively explored.

In this paper, a facile and efficient method is presented, where the *in situ* Ni-filled MWCNTs were synthesized with site-selectivity using chemical vapor deposition of propane on Si employing a simple mixture of Ni(salen) and a conventional photoresist as a modified photoresist (Mod-PR). Scanning electron microscopy (SEM), high-resolution transmission electron microscopy (HRTEM), energy dispersive x-ray (EDX) and Raman spectroscopy were used to characterize the morphology, internal structure and quality of the resultant products. The magnetization of the as-grown Ni-filled CNTs was studied as a function of the magnetic field using a superconducting quantum interference device (SQUID). We have also established a simple photolithographic process to fabricate catalyst patterns and Ni-filled CNTs have been synthesized over these lithographically defined catalyst regions, as observed via SEM analysis.

---
[4] Author to whom any correspondence should be addressed.

## 2. Experimental details

A Mod-PR solution of 0.2 M Ni(salen) concentration was prepared using a conventional positive photoresist as the dispersion medium (HPR 504, Fuji Film). HPR 504 is a positive photoresist which uses ethyl lactate as the solvent and has a viscosity of ∼40 cps. The solution was then stirred and sonicated for 30 min to achieve a good dispersion of the metal–organic molecular precursor. Then the solution was spin-coated with a rotation speed of 4000 rpm for 20 s on the Si(111) substrate to get a thin layer of the Mod-PR. The thin Mod-PR film was annealed in air for 1 h at 400 °C to improve the adhesion to the substrate. To the best of our knowledge, this is the first time Ni(salen) has been used as a catalyst for CNT growth.

The substrates were then loaded into a quartz tube furnace (Electroheat EN345T), pumped down to $10^{-2}$ Torr and backfilled with flowing argon to atmospheric pressure. Thereafter, the samples were heated in argon up to 900 °C following which the argon was replaced with hydrogen. Subsequently, the samples were annealed in a hydrogen atmosphere for 10 min. Finally, the reactor temperature was brought down to 850 °C and the hydrogen was turned off and propane introduced into the gas stream at a flow rate of 200 sccm for 1 h for CNT synthesis.

For lithographically selective growth of CNTs, after performing the spin-coating of Mod-PR the samples were baked at 90 °C for 15 min followed by an exposure step with a mask aligner to make an array of patterns. The exposed specimens were developed in the developer solution for 60 s and rinsed in distilled water. The synthesis of CNT was then carried out on the patterned areas following the same procedure as described earlier.

SEM (Zeiss SUPRA 40 and VEGA TESCAN) and HRTEM (JEOL JEM 2100) equipped with an EDX analyzer (Oxford Instruments) were employed for examination of the morphology and microstructure of the products. Raman measurements were carried out with a Renishaw RM1000B LRM at room temperature in the backscattering geometry using a 514.5 nm air-cooled Ar$^+$ laser as an excitation source for compositional analysis. The magnetic properties of the Ni-filled MWCNTs were measured using a SQUID at 5 K.

## 3. Results and discussion

The SEM image (figure 1) shows that the precursor powder, i.e. Ni(salen), comprises thin platelets with a broad range of crystallite sizes. Figure 2 shows the SEM image of the surface of the Mod-PR film over the Si(111) substrate after being annealed at 900 °C. The image shows that the catalyst nanoparticles are uniformly distributed on the Si substrate and their sizes range from about 10–100 nm.

The morphology of the as-synthesized carbon nanostructures was investigated by SEM. The SEM image (figure 3) shows the presence of a homogeneous distribution of CNTs covering the entire catalyst surface. From this image, it can be seen that most of the CNTs are nearly straight and long with a high number density. Furthermore, the wall surfaces of the CNTs appear relatively clean and smooth. EDX analysis of the as-grown CNTs (not shown here) shows that the material contains only carbon and Ni, with an Si peak (due to the substrate). The high magnification image of the CNTs (figure 4) shows metal particles at the tip of the nanotubes (marked in figure 4 with white arrows), indicating the tip-growth mechanism.

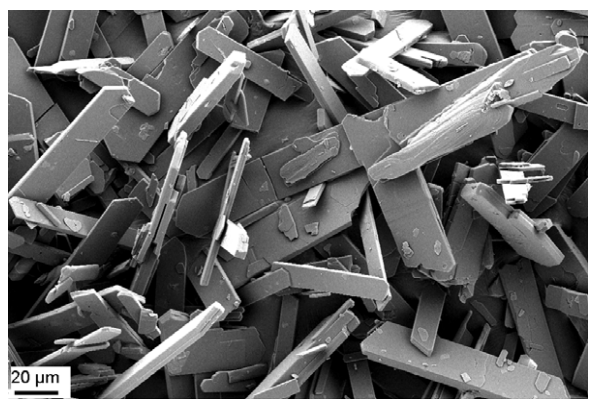

**Figure 1.** SEM image of the Ni(salen) powder.

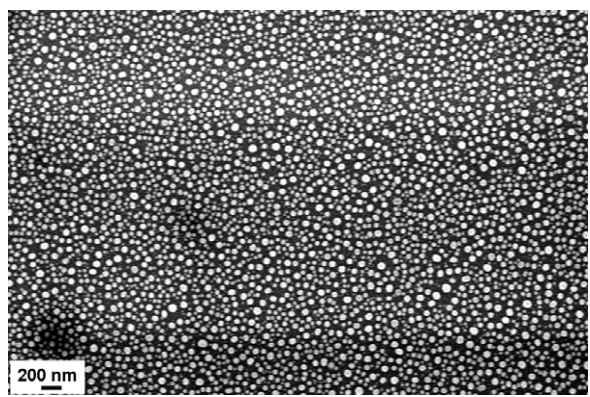

**Figure 2.** SEM micrograph of the catalytic nanoparticles prepared from Mod-PR film over the Si(111) substrate after annealing at 900 °C.

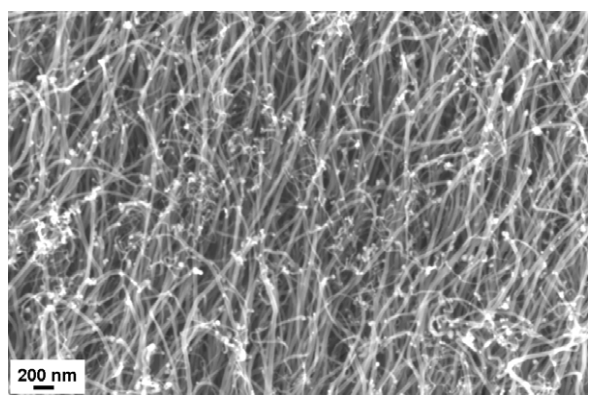

**Figure 3.** SEM micrograph of the MWCNTs synthesized using Mod-PR by CVD growth.

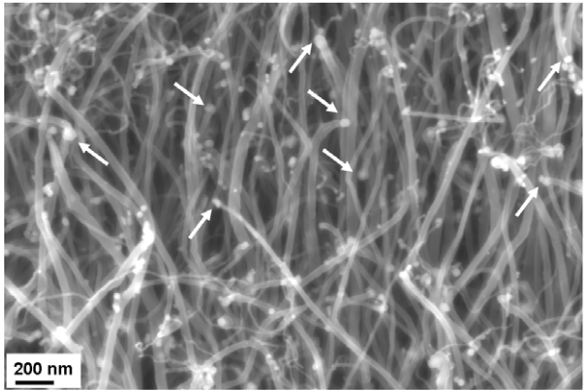

**Figure 4.** High magnification SEM micrograph of the MWCNTs synthesized using Mod-PR by CVD growth indicating the tip-growth mechanism.

In order to analyze the internal structure of the nanotubes more closely, HRTEM studies were performed. The sample preparation for HRTEM study was done by scraping the CNTs from the Si substrates and dispersing them ultrasonically in alcohol and transferring to carbon-coated copper grids. The TEM investigations reveal that the CNTs are the only product obtained and there are large numbers of long column-like metal nanowires in the cavities of CNTs (figure 5). All the nanowires are tightly wrapped by the nanotube wall, and their diameters vary with the inner diameter of the corresponding CNT (figures 5(a) and (b)). The length of the encapsulated nanowires (figure 5) is nearly hundreds of nanometers. The existence of the catalyst nanoparticles at the top of the CNTs suggest that the nanotubes grow via the tip-growth mechanism (figure 5(c)). There is almost no trace of metal nanoparticles on the outer surface of the nanotubes according to the TEM observations which indicate the high selectivity of the synthesis method in favor of metal-filled CNT formation.

Figure 6(a) is the TEM image of the CNT encapsulating a metal nanowire. It can be seen that the CNT is nearly straight with an outer diameter of about 65 nm and the length of the uniform metal nanowire encapsulated by the CNT is about 60 nm. Figures 6(b) and (c) are high magnification TEM images of the metal nanowire and the nanotube wall, respectively. The metal nanowire inside this nanotube exhibits clear well-ordered and straight lattice fringes (figure 6(b)).

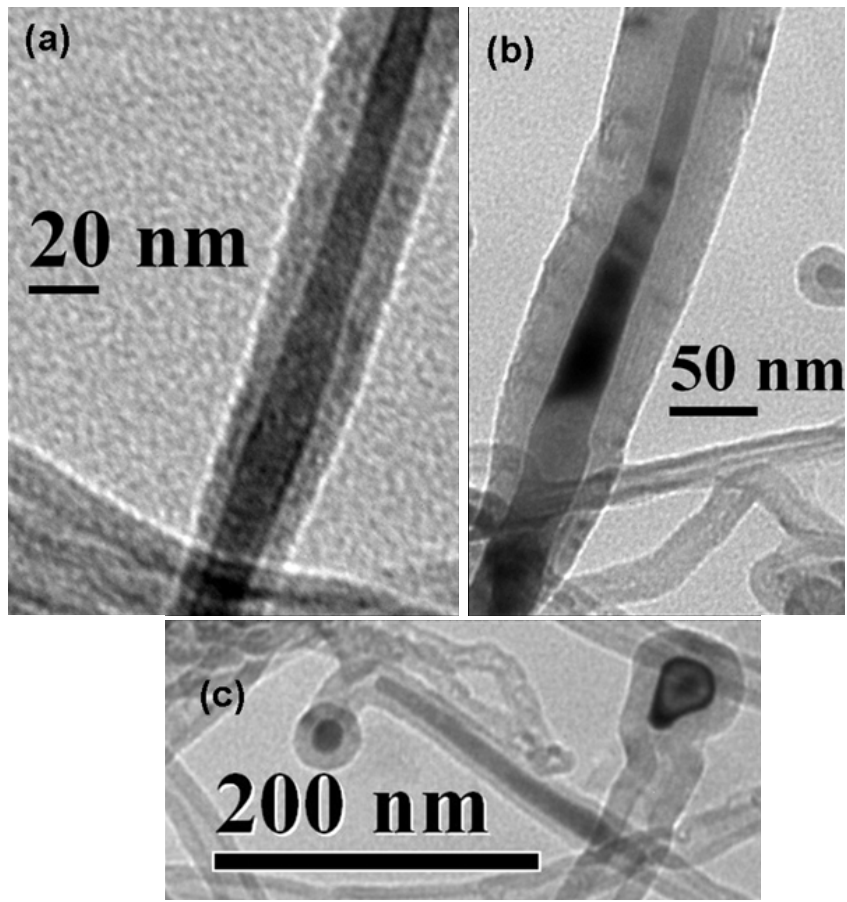

**Figure 5.** (a) and (b) TEM images of Ni-filled CNTs exhibiting continuous filling. (c) TEM images of as-formed MWCNTs indicating the tip-growth mechanism.

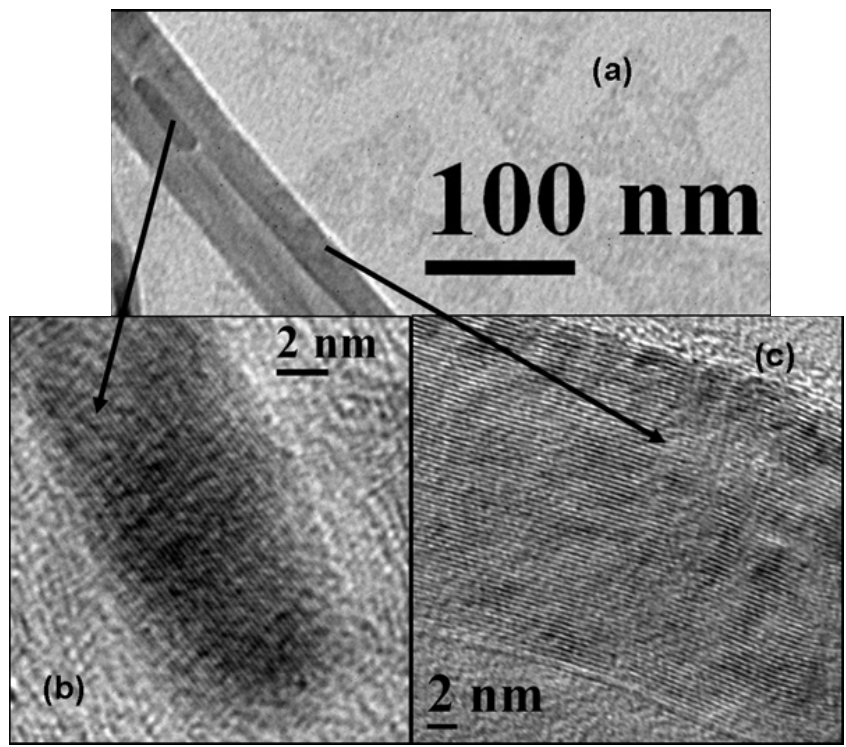

**Figure 6.** (a) TEM image of an as-formed MWCNT encapsulating a metal nanowire, (b) high-resolution TEM image of the CNT-encapsulated nanowire as indicated in (a). (c) High-resolution TEM image of the nanotube wall as indicated in (a).

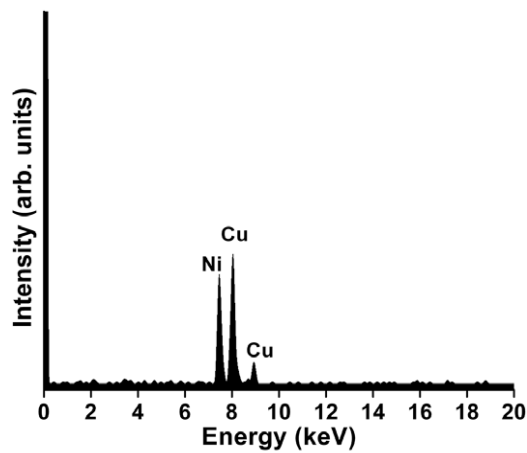

**Figure 7.** The EDX spectrum of the metal nanowire encapsulated within the CNT indicated in figure 6(a).

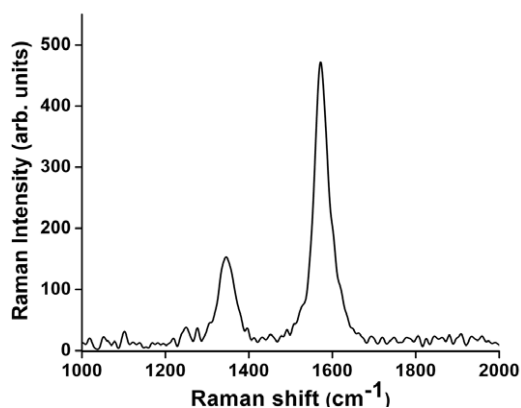

**Figure 8.** Raman spectrum (514.5 nm excitation) of the Ni-filled MWCNT film grown by CVD on Si using Mod-PR.

The $d$ spacings of the lattice fringes are 0.21 nm, which corresponds to the (111) crystal planes of Ni. Moreover, from figure 6(c), it is seen that the CNT is also well graphitized and the interplanar distance between two adjacent graphene planes is about 0.34 nm, which is close to that of the (002) interplanar distance in graphite, i.e. 0.335 nm. The higher interplanar distance observed in MWCNTs is attributed to the curved shape of their structures, which induce strain inside the graphene plane stacking [24].

The EDX analysis (figure 7) also demonstrates that the metallic nanowire inside the CNT indicated in figure 6(a) is of Ni, which is in agreement with the lattice spacing obtained from the HRTEM image (figure 6(b)). The Cu signals are due to the copper grid supporting the sample. The growth model of partially Ni-filled nanotubes can be explained by emphasizing the role of the capillary action of the liquid-like Ni particles that exist at the time of nanotube nucleation. The details of the growth model are discussed elsewhere [25].

Figure 8 shows the Raman spectrum for the as-grown CNTs at a laser excitation wavelength of 514.5 nm. The two main peaks are the D and G bands. The peak at 1572 cm$^{-1}$ (G band) corresponds to the E$_{2g}$ mode of graphite and is related to the vibration of sp$^2$-bonded carbon atoms, indicating the presence of crystalline graphite carbon in MWCNTs. The D band, at around 1346 cm$^{-1}$, is attributed to disordered the magnetization as a function of magnetic field of the as-prepared Ni-filled CNTs was measured at 5 K (figure 9) and it yielded a coercivity ($H_c$) of 94 Oe. This is significant in comparison to the bulk nickel value, where the coercive force is $H_c = 0.7$ Oe [27]. The coercivity of Ni-filled nanotubes is considerably enhanced as the magnetic properties of encapsulated Ni nanoparticles are closely associated with their size and internal structure, especially the shape anisotropy as mentioned before. More striking is the fact that the coercivity value (94 Oe) of partially Ni-filled MWCNTs in our case, is substantially higher than the previously reported low temperature ($T = 2$ K) coercivity value (40 Oe) of the same material [28]. Hence it may be concluded that the shape anisotropy of our Ni-filled nanotubes are significantly higher than previously reported values. This could have tremendous technological applications in magnetic recording technology.

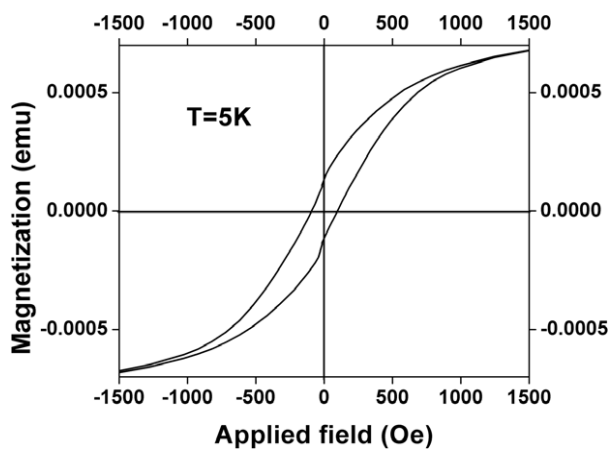

**Figure 9.** Magnetic hysteresis loop at $T = 5$ K of Ni-filled MWCNTs grown by CVD using Mod-PR.

graphitic materials. The intensity ratio of the D peak to the G peak, which measures quality, as derived from figure 8 is $R = I_D/I_G = 0.32$, indicating that the grown CNTs are highly crystalline in nature.

As the presence of shape anisotropy in a magnetic material can considerably enhance its magnetic properties [26], so the magnetic material with higher aspect ratio may have increased coercivity than its bulk form. To examine this,

The site-selective growth of CNTs over a pre-defined catalyst pattern was examined using SEM. Figure 10(a) is the SEM image of a Ni-containing patterns produced by the general photolithographic process using Mod-PR. Figure 10(b) is the SEM image of the grown CNTs on the catalyst pattern corresponding to figure 10(a), showing the high selectivity of the process. Figure 10(c) shows the SEM image of the site-selective growth of another carbon nanotube array, grown from lithographically defined catalyst lines and demonstrating the simplicity and scalability of the method.

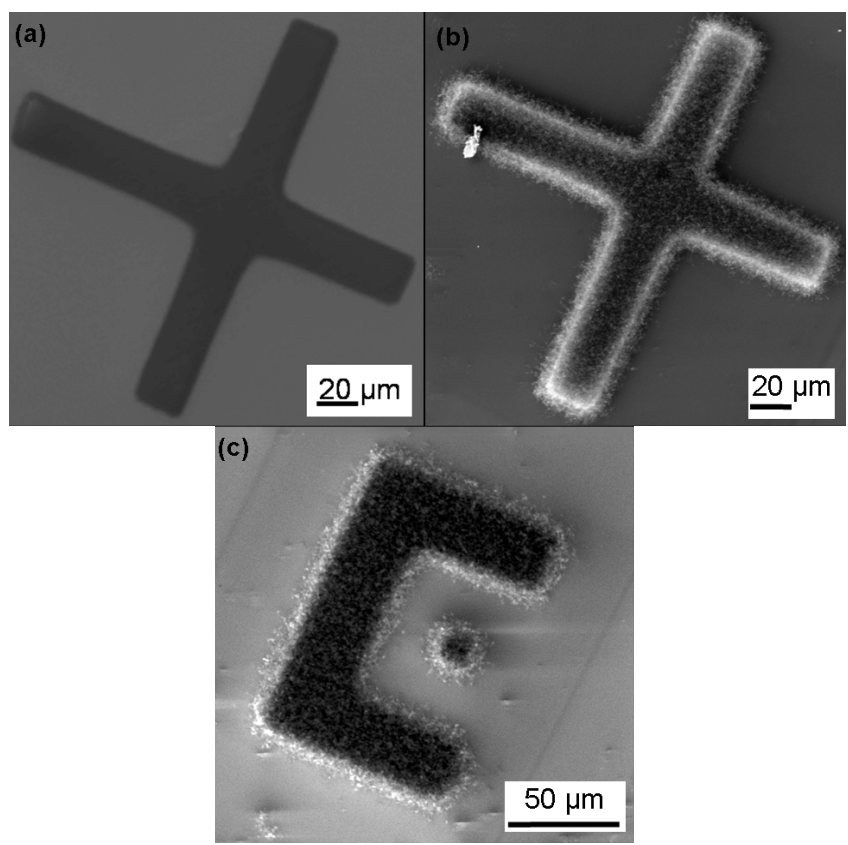

**Figure 10.** Site-selective growth of Ni-filled MWCNTs by direct photolithographic route using Mod-PR: (a) SEM image of a Mod-PR pattern, (b) SEM image of the Mod-PR pattern after CNT growth and (c) SEM image of another patterned growth of CNTs using Mod-PR.

## 4. Conclusions

A facile and efficient method has been demonstrated that can be used for simple and scalable growth of Ni-filled CNTs with site-selectivity using a mixture of Ni(salen) and conventional photoresist. The experimental results show that the Ni-filled CNTs are multi-walled, well graphitized and grown through a tip-growth mechanism. The Ni-filled MWCNTs show ferromagnetic behavior with large coercivity providing interesting possibilities for further applications in many potential areas, especially as magnetic recording media. Moreover, a feasible and effective photolithographic method has been devised for the spatially selective synthesis of high quality Ni-filled MWCNTs, which will grow only on a pre-defined catalyst surface. This photolithography-based approach is a simple and yet robust process with excellent potential to synthesize Ni-filled CNTs at lithographically defined regions in a reproducible fashion.

## Acknowledgment

We are grateful to Dr B Mishra from the Department of Geology and Geophysics, IIT Kharagpur for his help with the Raman measurement.